\documentclass[twocolumn,amsmath,floats,showpacs,nofootinbib]{revtex4}

\usepackage{graphicx}

\usepackage{textcomp}
\usepackage{hyperref}

\hypersetup{colorlinks=true}

\begin{document}

\title{Investigation of the electron-phonon interaction in $N{{b}_{3}}Sn$ with the aid of microcontacts}

\author{I. K. Yanson, N. L. Bobrov, L. F. Rybal'chenko and V. V. Fisun}

\affiliation
{B.I.~Verkin Institute for Low Temperature Physics and Engineering, of the National Academy of Sciences of Ukraine, prospekt Lenina, 47, Kharkov 61103, Ukraine\\
E-mail address: bobrov@ilt.kharkov.ua}
\published {Fiz. Tverd. Tela (Leningrad), Vol.27(6), p.1795 (1985); Soviet Physics Solid State, v. 27(6); p.1076 (1985)}
\date{\today}

\begin{abstract}The method of microcontact spectroscopy in the superconducting state was used to investigate weak nonlinearities of the current-voltage characteristics of point contacts made of $N{{b}_{3}}Sn$ single crystals. The nature of the spectrum of the electron-phonon interaction was found to vary considerably from contact to contact, indicating considerable deviations of the composition of the surface of $N{{b}_{3}}Sn$ from stoichiometry. A correlation was established between the nature of the spectrum and the magnitude of the gap singularities of the current-voltage characteristics. In the case of "dirty" high-resistance contacts with strong gap singularities the microcontact spectra were reasonably reproducible, which made it possible to relate them sufficiently closely to the microcontact function of the electron-phonon interaction in the bulk material. It was found that microcontact spectroscopy of this interaction was possible in the superconducting state not only in dirty $S-c-S$ contacts, but also in dirty $S-c-N$ contacts.
\end{abstract}
\maketitle

The phonon spectrum of intermetallics with the A-15 structure is exceptionally complex. It has been possible to reconstruct the dispersion curves from the experimental data along the principal crystallographic directions only in the case of $N{{b}_{3}}Sn$ (Ref. \cite{Pintschovius}), because this material can be obtained in the form of sufficiently large single crystals. In the case of $N{{b}_{3}}Sn$ only a calculation of the dispersion curves along the [111] direction is available \cite{Weber}. The phonon spectrum and particularly the electron-phonon interaction (EPI) in superconductors with the A-15 structure is very sensitive to deviations from the crystalline order and from stoichiometry. In this sense the EPI function of such materials is extremely labile. It has been pointed out in the literature \cite{Millier}, that there is a correlation between the displacement of the "center of gravity" of the EPI function in the direction of lower frequencies and the enhancement of superconducting properties.

We investigated the spectral functions of the EPI in $N{{b}_{3}}Sn$ obtained by differentiation of the low-temperature nonlinear current-voltage characteristics of microscopic electrical contacts made of this material. We selected the following pairs of electrodes: $N{{b}_{3}}Sn-Cu$, $N{{b}_{3}}Sn-N{{b}_{3}}Sn$, $N{{b}_{3}}Sn-Mo$. As demonstrated earlier, \cite{Yanson1} the second derivative of the current-voltage characteristic of a contact of the $N-c-S$ type ($N$ is a normal metal, $S$ is a superconductor, and $c$ is a constriction) is proportional to a linear combination of the microcontact EPI function of the electrodes on condition that $eV\gg \Delta $ and the contribution of each electrode is inversely proportional to the Fermi velocity. Since this velocity in $N{{b}_{3}}Sn$ is relatively low ($v_{F}^{N{{b}_{3}}Sn}\sim {{10}^{7}}cm/\sec$), we can expect predominance of the contribution of the relevant electrode to the characteristic observed in the case of heterocontacts. If contacts of the $S-c-S$ type are employed, the phonon singularities can be observed only in the case of complete destruction of the superconductivity in the constriction by some depairing centers \cite{Yanson2}.

Singularities of this type were observed by us for $N{{b}_{3}}Sn$. We found that the EPI functions of $N{{b}_{3}}Sn$ recorded by microcontacts have maxima which are in general agreement with the positions of the maxima found from the tunnel effect. The relative amplitudes of the maxima however, vary strongly from sample to sample, indicating a high sensitivity of the EPI to accidental changes in the structure and composition of the investigated material in a small region near a constriction. A correlation was found between the EPI function at low energies and the nonlinearity of the current-voltage characteristics in the range of displacement of the order of $\Delta$, reflecting the contribution of the superconducting state of $N{{b}_{3}}Sn$ to the EPI spectrum.

\section{PREPARATION METHOD}
The $N{{b}_{3}}Sn$  electrodes were single-crystal samples with a natural faceting and of $\sim 1\text{ }mm$ size. The absence of other phases from the bulk of a crystal was checked by the x-ray diffraction method. The degree of purity of the samples was characterized by the resistance ratio ${{{R}_{300}}}/{{{R}_{20.4}}\sim 5}\;$. Tbe crystal growth method was described in Ref. \cite{Matsakova}. Before mounting, these crystals were polished chemically in a mixture of concentrated acids ($HF+HN{{O}_{3}}+HCl{{O}_{4}}$) taken in the volume ratio 1:1:1; the polishing treatment lasted 30-40 sec and it was followed by careful washing in distilled water. After drying, the samples were soldered with indium to a special wire holder.

The traditional method \cite{Yanson3} was used to make electrodes of high-purity $Cu$ or $Mo$ (${{{R}_{300}}}/{{{R}_{4.2}}>1000}\;$). The sample was mounted in a pressure device and placed in a cryostat. Microcontacts were formed by the shear method and the force pressing the electrodes to the another was controlled outside the cryostat. The current-voltage characteristics and their derivatives were determined by the usual modulation method.
\begin{figure}[]
\includegraphics[width=8.7cm,angle=0]{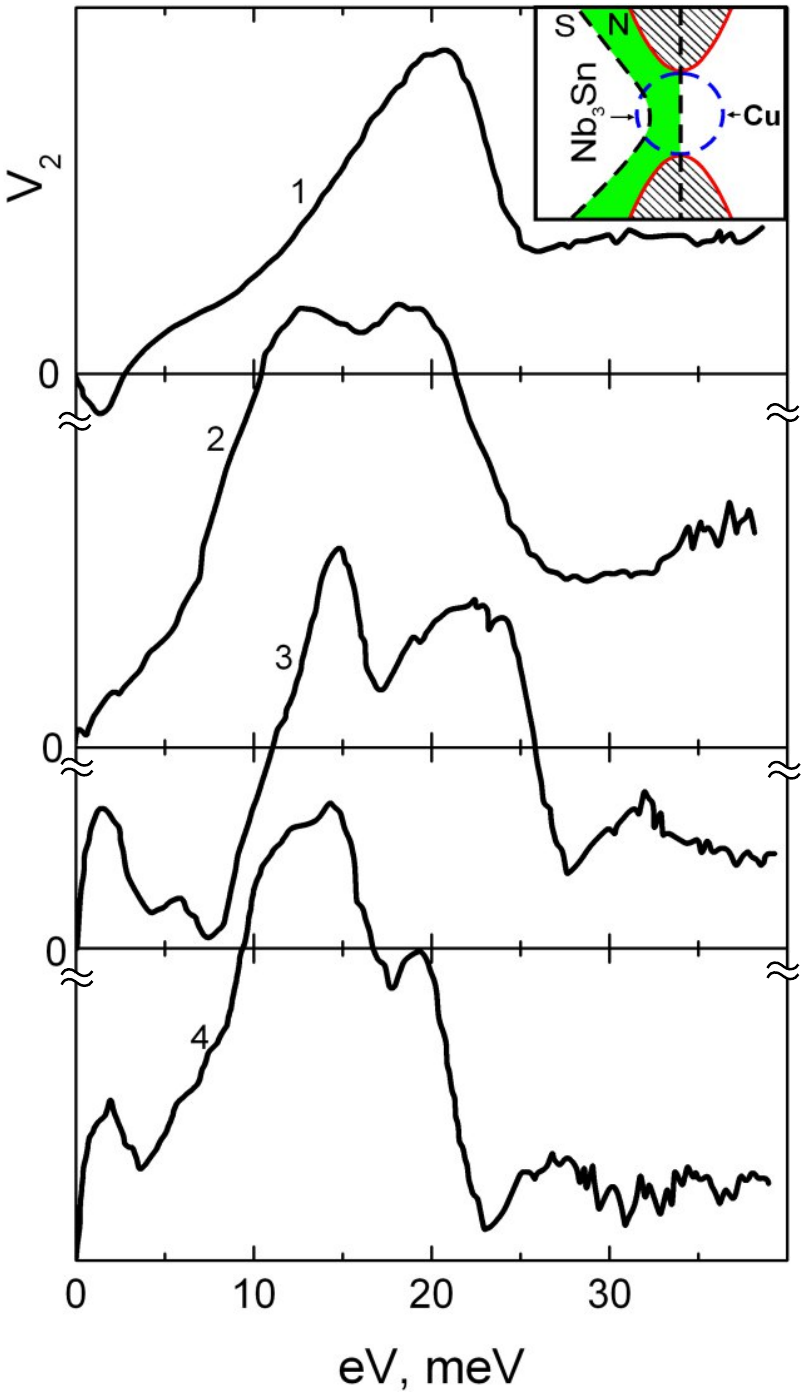}
\caption[]{Miciocontact spectra (${{V}_{2}}\sim{{d}^{2}}V/d{{I}^{2}}$) of pure  $N{{b}_{3}}Sn-Cu$ contacts $(H=0,\text{ }T=6.3-6.8\text{ }K$). Curves 1-4 represent contacts with resistances of 41, 388, 62, and 100 $\Omega $. The inset shows the proposed model of a contact.}
\label{Fig1}
\end{figure}

\begin{figure}[]
\includegraphics[width=8.5cm,angle=0]{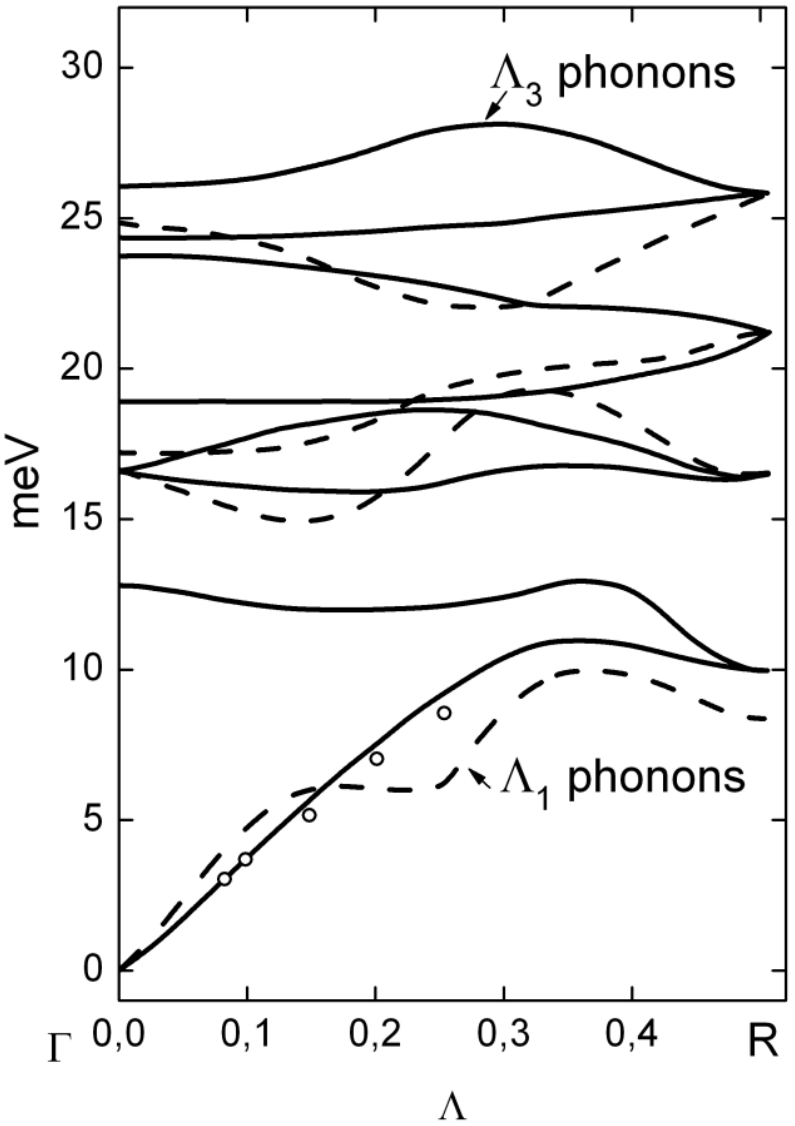}
\caption[]{Phonon dispersion curves $\omega (q)$ of $N{{b}_{3}}Sn$  in the [111] direction, found by calculation in Ref. \cite{Weber}}
\label{Fig2}
\end{figure}
\begin{figure}[]
\includegraphics[width=8.5cm,angle=0]{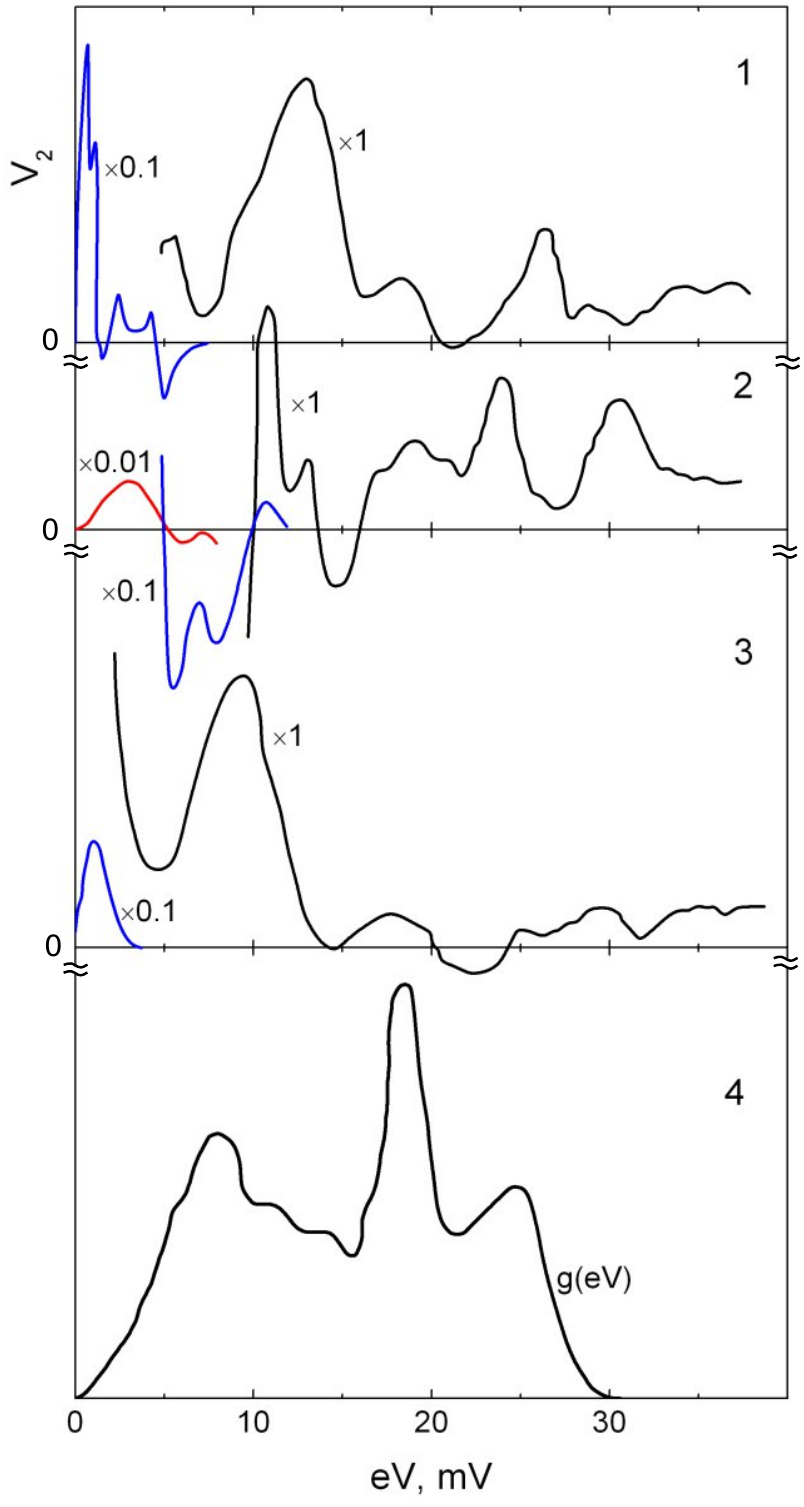}
\caption[]{Microcontact spectra of dirty contacts in zero magnetic field:\\1) $N{{b}_{3}}Sn-N{{b}_{3}}Sn$, $R=12\,\Omega ,\text{ }T\text{ }=\text{ }4.2\text{ }K;$
\\2) $N{{b}_{3}}Sn-Cu,\text{ }R=45\,\Omega ,\text{ }T\text{ }=\text{ }4.2\text{ }K;$
\\3) $N{{b}_{3}}Sn-Mo,\text{ }R=14\,\Omega ,\text{ }T\text{ }=\text{ }5\text{ }K;$
\\Curve 4 represents the tunnel function of the EPI in $N{{b}_{3}}Sn$ (Ref. \cite{Vedeneev2}).
}
\label{Fig3}
\end{figure}

\section{EXPERIMENTAL RESULTS AND DISCUSSION}
In studies of nonlinear phenomena in the electrical conduction of metallic microcontacts it is always necessary to take measures against the stray influence of the Joule heating of the metal In the region where the current is concentrated and this effect becomes significant when the contact diameter is increased by a value of the order of the inelastic electron relaxation length ${{\Lambda }_{\varepsilon }}$. In the microcontact spectroscopy method it is therefore preferable to study the EPI employing contacts with, if possible, large resistances. Since such contacts have automatically the smallest dimensions, the requirements in respect of the purity and structural perfection of the surface layer of the electrodes become extremely onerous. However, it is not always possible to satisfy in practice these requirements in respect of the $N{{b}_{3}}Sn$ electrodes.

Deviations from stoichiometry and crystalline order are obviously the cause of considerable variations of the intensities of the peaks in the microcontact spectra (which represent the dependences of ${{V}_{2}}\sim {{d}^{2}}V/d{{I}^{2}}$ on $V$) of the high-resistance $N{{b}_{3}}Sn-Cu$ contacts shown in Fig. \ref{Fig1}. All the spectra show a positive quantity (representing the voltage of the second harmonic of the modulating signal), which indicates that they are in agreement with the theory of Khlus \cite{Khlus} on condition that the contribution due to the same dissipation processes as in the normal state predominates. In other words, the nonlinearity due to the energy dependence of the excess current is unimportant. The spectral bands occupy the energy range up to 27-30 $meV$, which is in agreement with the calculated results presented in Fig. \ref{Fig2}. The absence of bands at higher energies shows that the observed peaks are indeed due to the EPI and not due to any stray factors (such as destruction of the superconductivity).

It is known \cite{Yanson3} that copper has its main band at $eV\sim 16-20\text{ }meV$. This band is missing from the spectra in Fig. \ref{Fig1}. Consequently, the recorded spectra apply to  , which is to be expected because ${v_{F}^{N{{b}_{3}}Sn}}/{v_{F}^{Cu}\ll 1}\;$. We must also draw attention to the fact that the background beyond the limits of the phonon spectrum is small for the selected curves. This provides an indirect evidence of the absence of the Joule heating for the samples in question because in the thermal limit the background level is high at $eV\gg \hbar {{\omega }_{\max }}$. A maximum at $eV\sim{\ }2-3\text{ }meV$ exhibited by some of the curves in Fig. \ref{Fig1} clearly corresponds to the energy in deeper superconducting $N{{b}_{3}}Sn$ layers (inset in Fig. 1). Its intensity is proportional to the contribution of the microcontact EPI function of the superconducting modification of $N{{b}_{3}}Sn$ to the observed microcontact spectrum, which represents the EPI characteristic averaged over the whole range of concentrations of the current. Clearly, the higher the intensity of this maximum, the greater the intensity of the low-frequency bands in the EPI spectrum. For example, spectrum 1 in Fig. \ref{Fig1}, which is entirely due to the normal layer on the surface of $N{{b}_{3}}Sn$, not only does not have a low-energy maximum, but even exhibits a negative "zero-point anomaly," which is usual in the case of normal metals \cite{Yanson3}. In this case the EPI spectrum is concentrated entirely at high phonon energies.

The example of $Nb$ was used by us in Ref. \cite{Yanson2} to show that the microcontact spectroscopy method can be applied to the superconducting state if the superconductivity in the contact region is greatly weakened, but the thermal effect is still quite small. This situation arises if some depairing centers are introduced into the region of the constriction in the course of electrochemical preparation of the electrodes. The spectrum may then be displacec partly to the negative range. Therefore, in such "dirty" contacts we can expect a considerable enhancement (compared with the $N$ state) of the phonon singularities and their intensities are then almost entirely due to the energy dependences of the excess current.

The microcontact spectrum of such a dirty $N{{b}_{3}}Sn-N{{b}_{3}}Sn$ contact is represented by curve 1 in Fig. \ref{Fig3}. The superconductivity of this contact is suppressed only near the center of the contact, so that in spite of the absence of the dc Josephson effect and in spite of the smallness of the excess current, the intensity of the gap singularity is particularly large. The shift of the spectrum toward lower frequencies is associated with the latter circumstance. It should be pointed out that similar spectra have been observed earlier in the $N{{b}_{3}}Sn-I-Cu$ tunnel contacts with microscopic short circuits \cite{Vedeneev1}.

In many investigated $N{{b}_{3}}Sn-Cu$ contacts the current flow mechanism does not correspond to the lower limit, as indicated by the low relative magnitude of the excess current which does not exceed the, theoretical value for dirty contacts. This cannot be associated with the suppression of the superconductivity at the edges of the contact, because the gap singularity in the contacts under discussion has a large amplitude. Moreover, the form of the spectrum is close to that observed for dirty $S-c-S$ contacts. A typical spectrum of a dirty $N{{b}_{3}}Sn-Cu$ contact is shown in Fig. \ref{Fig3} (curve 2). The same figure also shows that a change of the metal used for the normal electrode does not alter the spectrum (curve 3).

One of the possible factors which make the microcontact spectroscopy of the EPI possible in the case of dirty $S-c-N$  contacts may be similar to that suggested by us earlier \cite{Yanson2} for $S-c-S$ contacts. Its essence is as follows: in the case of contacts with a spatially inhomogeneous distribution of impurities in the superconducting electrode near a constriction the electrochemical potential of the Cooper pairs falls at distances shorter from the center of the contact than the chemical potential of the normal quasiparticles. This ensures the presence of a steep step of magnitude $eV$ exhibited by the nonequilibrium distribution function near the center of the contact, known to be the necessary condition for the microcontact spectroscopy method \cite{Yanson3}. According to this hypothesis, only the region of the contact on the superconducting electrode side may be dirty, whereas the second (normal) electrode should then remain pure. Moreover, we cannot exclude the possibility that a defect layer with a low transparency may form at the boundaries of the electrodes in the process of preparation and that electron tunneling takes place across the layer. In this case the appearance of the phonon singularities in the current-voltage characteristics or their derivatives in the superconducting state may be partly or completely due to the tunnel effect.

The microcontact spectra shown in Fig. \ref{Fig3} are characterized by a much higher reproducibility in respect of the positions of the phonon singularities than those shown in Fig. \ref{Fig1}. Some variations among these more reproducible spectra (curves 1-3 in Fig. \ref{Fig3}) are obviously due to small deviations from the stoichiometry on the surface of the $N{{b}_{3}}Sn$ electrodes. A comparison with the tunnel results \cite{Vedeneev2} (curve 4 in Fig. \ref{Fig3}) shows that the microcontact spectra exhibit particularly clearly the low-frequency singularities and this is obviously due to the energy dependence of the $K$ factor. (In a rough approximation lor tne $N-c-N$ contact the microcontact function is divided by the $Ê$ factor to obtain the equilibrium Eliashberg function which describes very closely the tunnel spectrum of the EPI. In the case of the microcontact spectra of $N{{b}_{3}}Sn$ this procedure cannot be applied because there is as yet no theoretical basis for the dirty $S-c-N$ contacts.) The greatest similarity between the positions of the phonon singularities and the tunnel data is exhibited by the microcontact spectrum of the $N{{b}_{3}}Sn-Cu$  system (curve 2 in Fig. \ref{Fig3}), although in the case of the latter there is an additional high-frequency peak near the energy $\sim 30\text{ }meV$, which however does not contradict the nature of the dispersion curves (Fig. \ref{Fig2}). It is necessary to stress that the tunnel results are obtained for film structures, in which we cannot avoid interlayer contamination and which has a considerable influence on the nature of the tunnel spectra. In view of the fact that pure single-crystal $N{{b}_{3}}Sn$  electrodes were used in our experiments we may assume that the microcontact spectra (Fig. \ref{Fig3}) represent more closely the real characteristics of the EPI in bulk $N{{b}_{3}}Sn$ than do the tunnel spectra.

The curves in Fig. \ref{Fig1} represent the contacts which include superconducting regions with a considerable deviation from stoichiometry accompanied by a fairly strong weakening of the superconductivity. If we bear in mind that the ballistic flow of the current takes place in these contacts, we can postulate also that the nonsuperconducting (at $4.2\text{ }K$) phases $N{{b}_{6}}S{{n}_{5}}$  or $NbS{{n}_{2}}$ are also present in the constriction.

Consequently, the extreme lability of the EPI of this material makes it very difficult to obtain reproducible characteristics corresponding to the properties of the bulk superconductor. Moreover, the anisotropic properties of the crystal structure of $N{{b}_{3}}Sn$ can also have some effect.

The authors are grateful to B. G. Lazarev and A. A. Matsakova for supplying   single crystals.

\end{document}